\documentclass[12pt]{iopart}
\expandafter\let\csname equation*\endcsname\relax
\expandafter\let\csname endequation*\endcsname\relax
\usepackage[LGR,T1]{fontenc}
\usepackage[utf8]{inputenc}
\usepackage{textcomp}
\usepackage{amsmath}
\usepackage{amssymb}
\usepackage{graphicx}
\usepackage{esint}
\usepackage{subscript}
\usepackage{cite}

\begin{document}
\title[All-optical]{All-optical three-dimensional electron pulse compression}

\author{Liang Jie Wong$^1$$^,$$^3$, Byron Freelon$^2$,
Timm Rohwer$^2$, Nuh Gedik$^2$, and
Steven G Johnson$^1$}

\address{$^1$Department of Mathematics, Massachusetts Institute
of Technology, 77 Massachusetts Avenue, Cambridge, Massachusetts 02139,
USA}
\address{$^2$Department of Physics, Massachusetts Institute
of Technology, 77 Massachusetts Avenue, Cambridge, Massachusetts 02139,
USA}
\address{$^3$Current Address: Singapore Institute of Manufacturing Technology, 71 Nanyang Drive, Singapore 638075}
\ead{ljwong@alum.mit.edu}

\begin{abstract}
We propose an all-optical, three-dimensional electron pulse compression
scheme in which Hermite-Gaussian optical
modes are used to fashion a three-dimensional optical trap in
the electron pulse's rest frame. We show that the correct choices
of optical incidence angles are necessary for optimal compression.
We obtain analytical expressions for the net impulse imparted by Hermite-Gaussian
free-space modes of arbitrary order. Although we focus on electrons,
our theory applies to any charged particle and any particle with non-zero
polarizability in the Rayleigh regime. We verify our theory numerically
using exact solutions to Maxwell's equations for first-order Hermite-Gaussian
beams, demonstrating single-electron pulse compression factors of
$>10^{2}$ in both longitudinal and transverse dimensions with experimentally
realizable optical pulses. The proposed scheme is useful in ultrafast
electron imaging for both single- and multi-electron pulse compression,
and as a means of circumventing temporal distortions in magnetic lenses
when focusing ultrashort electron pulses. Other applications include the creation of flat electron beams and ultrashort electron bunches for coherent terahertz emission.
\end{abstract}



\maketitle

\section{Introduction}

The ability of ultrafast X-ray and electron pulses to probe structural
dynamics with atomic spatiotemporal resolution has fueled a wealth
of exciting research on the frontiers of physics, chemistry, biology
and materials science \cite{Emma2010,Mancuso2010,Sciani2011,Zewail2006}.
Although electrons lack the penetration depth of X-rays, the large
scattering cross section of electrons (10\textsuperscript{5}-10\textsuperscript{6}
times that of X-rays of the same energy \cite{Dwyer2006,Carbone2012})
and relative availability of high intensity table-top electron sources
favor the use of electrons especially in the study of surfaces, gas
phase systems and nanostructures.

An electron pulse tends to expand and acquire a velocity chirp as
it travels, firstly due to space-charge (i.e. inter-electron repulsion),
and secondly due to dispersion resulting from an initial velocity
spread. The propagation of electron pulses has been the subject of extensive study \cite{Siwick2002,Michalik2006,Reed2006}. To ensure that the electron pulse arrives at the sample or detector with the desired properties (e.g., spot size, coherence length,
pulse duration), many ultrafast electron imaging setups adopt means
to compress the electron pulse transversely and longitudinally. Longitudinal
compression methods include the use of electrostatic elements \cite{Wang2012},
microwave cavities \cite{Gao2012,vanOudheusden2010,Kassier2012,Chatelain2012,Gliserin2012}
and optical transients \cite{Baum2007,Hilbert2009}. These techniques
can potentially compress single-electron pulses \cite{Aidelsburger2010,Zewail2010}
to attosecond-scale durations \cite{Baum2007,Veisz2007}. Tranverse
compression, or focusing, of an electron pulse is typically achieved
with standard charged particle optics like magnetic solenoid lenses.
Femtosecond electron pulses, however, suffer significantly from temporal
distortions in magnetic lenses and require more complicated combinations
of charged particle optics for isochronic imaging \cite{Weninger2012}.

In this paper, we propose a scheme for the three-dimensional compression
of electron pulses using only optical transients, with no static fields
involved. The scheme comprises a succession of Hermite-Gaussian optical
modes that effectively fashions a three-dimensional optical trap in
the electron pulse's rest frame. Such a scheme is useful in ultrafast
electron imaging for both single- and multi-electron pulse compression,
and as a means of circumventing temporal distortions in magnetic lenses
\cite{Weninger2012} when focusing ultrashort electron pulses. Methods
of generating Hermite-Gaussian modes include the use of waveplates
\cite{Novotny1998} and excitation in diode lasers \cite{Chu2012}.

In Section 2, we present an overview of the three-dimensional electron
pulse compression scheme, and describe how a succession of compression
stages may be implemented with a single optical pulse. In Section
3, we show mathematically that the right choice of optical incidence
angle is necessary for optimal longitudinal compression, and obtain
analytical expressions for the net velocity change induced in a charged
particle by the passage of an optical pulse. In Section 4, we illustrate
the conclusions of Section 3 with exact numerical simulations of the
laser-electron interaction. We demonstrate single-electron pulse compression
factors of $>10^{2}$ in both longitudinal and transverse dimensions
using experimentally-realizable optical pulses, and study the energy
scaling laws of the compression scheme.

\section{Overview}

A charged particle in an electromagnetic wave experiences a time-averaged
force called the ponderomotive force \cite{Eberly1991,Boot1957} that
pushes the particle towards regions of lower optical intensity in
the particle's rest frame. Dielectric particles are also subject to
this phenomenon, and applications of electromagnetic ponderomotive
forces have included atomic cooling, optical manipulation of living
organisms, plasma confinement, and electron acceleration \cite{Ashkin1997,Dodin2007,Stupakov2001}. The optical ponderomotive force has also been used in the characterization of ultrashort electron pulses \cite{Siwick2005,Hebeisen2006,Morrison2013,Hebeisen2008}.

\begin{figure}
\includegraphics[width=12cm]{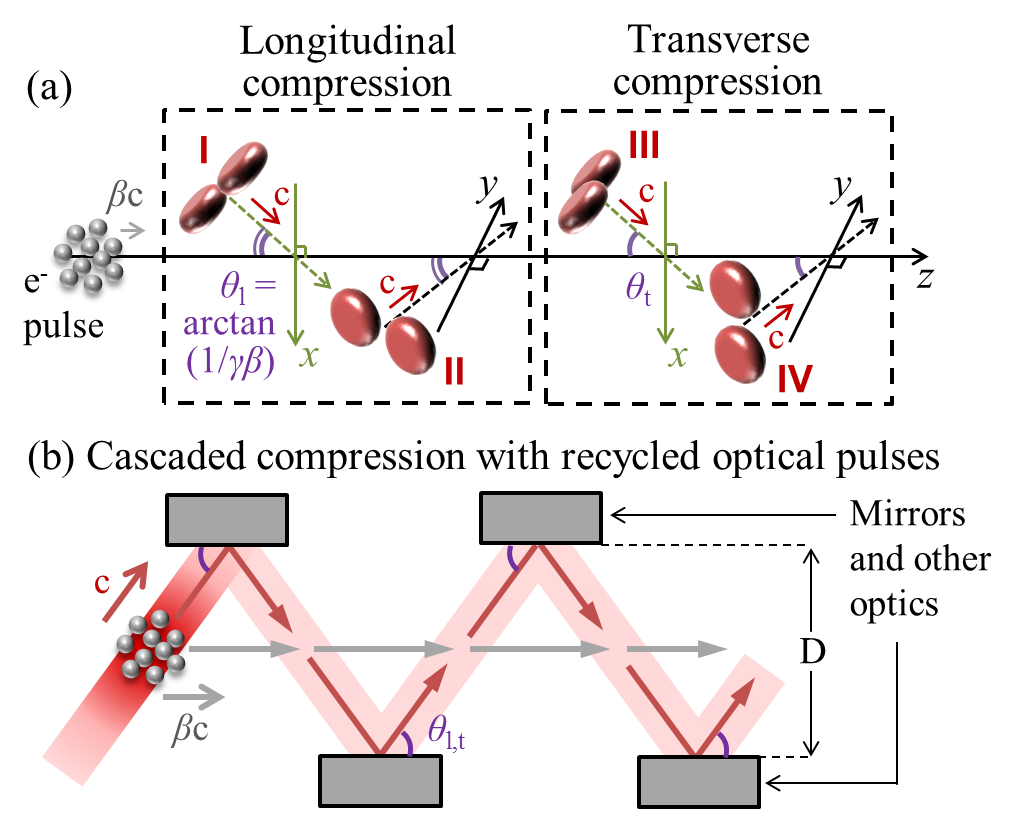}\caption{(a) Schematic diagram of three-dimensional electron pulse compression
technique using pulsed first-order Hermite-Gaussian optical modes,
which are portrayed as pairs of shiny red lozenges. Green lines lie
in the x-z plane, black lines in the y-z plane. Dotted lines are the
beam axes down which the optical pulses propagate. The electron pulse
travels at speed $v\equiv\beta\textnormal{c}$ in the +z-direction,
c being the speed of light in vacuum. $\gamma\equiv\left(1-\beta^{2}\right)^{-1/2}$
is the Lorentz factor. (b) Schematic diagram illustrating how a single
optical pulse may be used to implement a succession of compression
stages. Lines ending in filled arrowheads sketch the trajectories
of optical (red) and electron (gray) pulses, with the arrowheads terminating
at the interaction points.\label{Fig:Schematic}}
\end{figure}

Here, we use the ponderomotive force to compress an electron pulse
by subjecting the electron pulse to the intensity minimums of appropriately-oriented
Hermite-Gaussian modes, as illustrated in Fig. \ref{Fig:Schematic}(a).
Compression in each Cartesian dimension can be controlled without
affecting electron pulse properties in orthogonal dimensions, at the
lowest order. Although either Pulse I or II suffices for longitudinal
compression, using two identical pulses in the configuration shown
ensures that any higher-order modulations affecting transverse electron
pulse properties do so equally in x and y. In Fig. \ref{Fig:Schematic}(a),
Pulses I and II control compression in z, whereas Pulses III and IV
control compression in y and x respectively. The stages (and optical
pulses) may be arbitrarily ordered and cascaded, as long as inter-particle
interactions and dispersion affect the electron pulse negligibly between
interactions. Since the ponderomotive force is a non-linear effect
(i.e. not directly proportional to electric field), the optical pulses
should be sufficiently far apart so that interference between the
fields of different pulses does not occur.

The use of an optical pulse's transverse intensity profile for electron
pulse compression has been proposed in \cite{Hilbert2009}. However,
the scheme in \cite{Hilbert2009} uses an optical incidence angle
normal to the electron path in the lab frame, a sub-optimal configuration
for electrons of non-zero speed. In addition, the scheme in \cite{Hilbert2009}
uses a Laguerre-Gaussian ``donut'' mode, which -- even for a stationary
electron pulse -- couples compression in the longitudinal dimension
to that in exactly one transverse dimension.

\begin{figure}
\includegraphics[width=12cm]{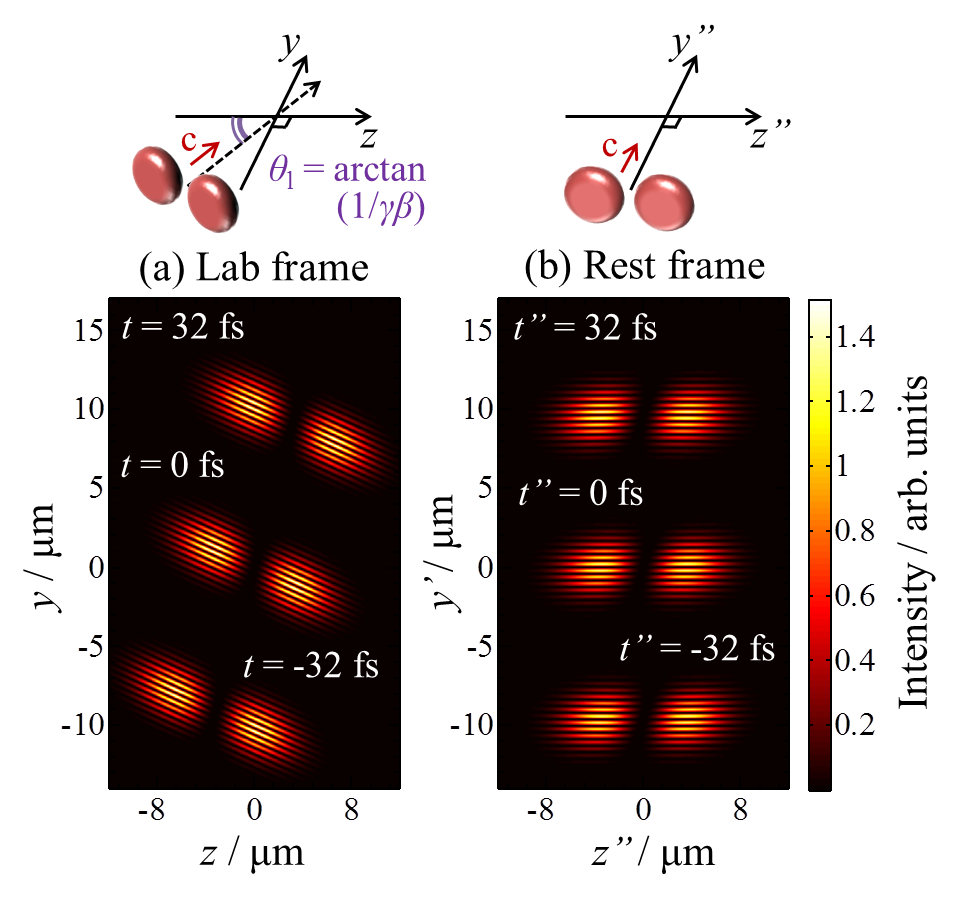}
\caption{Intensity profile of Pulse II (c.f. Fig. 1(a)) at three instances
in time in the (a) lab frame and (b) rest frame of a 30 keV electron
pulse. In the lab frame, the temporal pulse and carrier wavefront
are obliquely incident at $\theta_{\textnormal{l}}=70.9^{0}$, in
accordance with (\ref{best theta}), giving rise to normal incidence
in the rest frame. Double-primes denote rest frame variables. \label{Fig:LabRestFrame}}
\end{figure}

Intuitively, the oblique optical incidence angle of the longitudinal
compression stage is motivated by the desire for normal optical incidence
in the electron pulse's rest frame. This implies a lab frame incidence
angle of
\begin{equation}
\theta_{\textnormal{l}} =\arctan\left[\frac{\textnormal{c}\sin\theta_{\textnormal{l}}''}{\gamma\left(\textnormal{c\ensuremath{\cos\theta_{\textnormal{l}}''}}+v\right)}\right]= \arctan\left(\frac{1}{\gamma\beta}\right),\label{best theta}
\end{equation}
where the electron pulse propagates in the +z-direction with speed
$v\equiv\beta\textnormal{c}$ (c the speed of light in vacuum), corresponding
to Lorentz factor $\gamma\equiv\left(1-\beta^{2}\right)^{-1/2}$.
The first equality in (\ref{best theta}) expresses the relation between
the rest frame incidence angle $\theta_{\textnormal{l}}''$ and $\theta_{\textnormal{l}}$
. The second equality was made by setting $\theta_{\textnormal{l}}''=90^{\textnormal{o}}$.
We have taken the optical group velocity as c, a valid assumption
\cite{Esarey1995} for the paraxial, many-cycle optical pulses we
are interested in. The physics behind (\ref{best theta}) is illustrated
in Fig. \ref{Fig:LabRestFrame}, which shows how oblique optical incidence
in the lab frame corresponds to normal optical incidence in the electron
pulse's rest frame. In the next section, we show mathematically that (\ref{best theta}) is optimal in the sense that when it is satisfied, the induced velocity change in the longitudinal direction is not a function of transverse coordinates and not accompanied by transverse
phase plane modulations, at the lowest order. Fig. \ref{Fig:DifferentExamples}(a) illustrates the physical mechanism of the longitudinal compression scheme: the laser-electron interaction induces a velocity modulation in the electron pulse, which then compresses as it continues to propagate. The transverse compression scheme works according to the same principles, except that the desired velocity modulation is now along a transverse dimension.

Since the electron pulse is stationary along its transverse dimensions,
normal incidence in the rest frame is achieved with any value of $\theta_{\textnormal{t}}$
for transverse compression. Indeed, we see in Section 4 that the transverse
compression of 30 keV electrons is a relatively weak function of $\theta_{\textnormal{t}}$.
However, the choice of $\theta_{\textnormal{t}}$ can significantly
affect the longitudinal compression ratio in a three-dimensional compression
scheme via higher-order terms of the transverse compression stage,
with the best results achieved when $\theta_{\textnormal{t}}=0^{\textnormal{o}}$.

Equation (\ref{best theta}) is also the condition for group velocity matching
between electron and optical pulses along the axis of electron pulse
propagation (i.e. $\textnormal{c\ensuremath{\cos}}\theta_{\textnormal{l}}=\textnormal{c}\beta\equiv v$).
This observation motivates the cascaded compression scheme of Fig.
\ref{Fig:Schematic}(b), in which an optical pulse (either Pulse I,
II, III or IV) is reflected and re-focused by a succession of optical
stages, so as to be repeatedly incident upon the electron pulse, allowing
the optical pulse to be utilized to its maximum capacity. If (\ref{best theta})
is satisfied, the interval between laser-electron coincidences is
\begin{equation}
T_{\textnormal{coin}}=\frac{\gamma D}{\textnormal{c}},\label{Tcoin}
\end{equation}
assuming that the electron pulse is injected along the axis of symmetry
of the setup, and that the optical components introduce no delays.
To avoid optical interference between successive interactions, $D$
should generally be chosen so that $T_{\textnormal{coin}}\gg\tau$
is satisfied, $\tau$ being the optical pulse duration. With suitable
combinations of optics, one can also implement the design in Fig.
\ref{Fig:Schematic}(b) for any optical incidence angle, or such that
a single optical pulse is used to realize several or all of Pulses
I, II, III and IV, since the four types of pulses essentially differ
only in orientation.

\section{Theory}

In this section, we obtain analytical expressions approximating the
ponderomotive potential and net impulse transfer associated with transverse
and longitudinal compression by pulsed Hermite-Gaussian $\textnormal{TEM}{}_{mn}$
modes of arbitrary order. We show mathematically that when (\ref{best theta})
is satisfied, the induced velocity change for longitudinal compression
is not a function of transverse coordinates and not accompanied by
transverse phase plane modulations, at the lowest order. Although
we focus on charged particles, our treatment may be extended to any particle
with non-zero polarizability in the Rayleigh regime (particle size
much smaller than electromagnetic wavelength) by the simple replacement
of a constant factor.

A charged particle in an electromagnetic wave experiences a force
\cite{Eberly1991,Boot1957}
\begin{equation}
\vec{F}=-\nabla U_{\textrm{p}}+...,\label{force}
\end{equation}
where the ponderomotive potential $U_{\textnormal{p}}$ is 
\begin{equation}
U_{\textrm{p}}\equiv\frac{q^{2}}{4m_{0}\omega^{2}}\left|\vec{E}_{a}\right|^{2},\label{Up}
\end{equation}
and $q$ and $m_{0}$ are respectively the particle's charge and rest
mass. The particle sees the electric field $\vec{E}=\left(\vec{E}_{a}\textrm{e}^{\textrm{i}\omega t}+\textrm{c.c.}\right)/2$,
where $\vec{E}_{a}$ varies slowly in time compared to the carrier
factor and $\textrm{i}\equiv\sqrt{-1}$. The ellipsis in (\ref{force})
hides terms proportional to $\textrm{e}^{\textrm{\ensuremath{\pm}i}\omega t}$
or $\textrm{e}^{\textrm{\ensuremath{\pm}i2}\omega t}$. Equation (\ref{force})
was derived from the Newton-Lorentz equation in the rest frame of
the initial particle. As such, the notion that a particle experiences
a force proportional to the gradient of electromagnetic intensity
is valid in the rest frame of the particle, and not necessarily in
a frame where the particle moves with any substantial velocity. The
net momentum imparted to a particle by the passage of a many-cycle
pulse is then
\begin{equation}
\triangle\vec{p}=\int\vec{F}\textrm{d}t=-\int\nabla U_{\textrm{p}}\textrm{d}t,\label{dp}
\end{equation}

Physically, the electric field causes the charged particle to oscillate
about its initial position, generating an effective dipole that is
subject to the same radiation pressure forces \cite{Harada1996} experienced
by dielectric particles in optical tweezers \cite{Ashkin1997}. In
fact, replacing $q/m_{0}\omega^{2}$ by $\alpha/2$ turns (\ref{Up})
into the ponderomotive potential of a particle in the Rayleigh regime,
where the particle's polarization $\vec{P}=\alpha\vec{E}$. The results
in this paper thus also apply to polarizable particles.

A paraxial, many-cycle electromagnetic pulse can be modeled using
the vector potential ansatz
\begin{equation}
\vec{A}=\textrm{Re}\left\{ \vec{\tilde{A}}\textrm{e}^{\textrm{i}\psi}g\left(\frac{\xi}{\xi_{0}}\right)\right\} ,\label{A}
\end{equation}
where each component of $\vec{\tilde{A}}$ is a solution of the paraxial
wave equation \cite{Mandel1995}, $g\left(\cdot\right)$ a real even
function describing the pulse shape such that $\underset{\left|\xi\right|\rightarrow\infty}{\lim}g\left(\xi\right)\rightarrow0$,
$\xi_{0}$ a constant associated with pulse duration, $\xi\equiv\omega t-k(z-z_{\textrm{i}})$
and $\psi\equiv\xi+\psi_{0}$, with $z_{\textrm{i}}$ the pulse's
initial position (at $t=0$) and $\psi_{0}$ a phase constant. $x$,
$y$ and $z$ are Cartesian coordinates. $\vec{\tilde{A}}$ is a slowly-varying
function of only spatial coordinates such that $\partial_{\textrm{x}}\vec{\tilde{A}},\,\partial_{y}\vec{\tilde{A}}=\textrm{O}\left(\epsilon_{\textrm{d}}\right)$
and $\partial_{\textrm{z}}\vec{\tilde{A}}=\textnormal{O}\left(\epsilon_{\textrm{d}}^{2}\right)$,
where the beam divergence angle $\epsilon_{\textrm{d}}\ll1$. To ensure
that the particle bunch interacts with the electromagnetic pulse only
when the bunch is close to the electromagnetic beam axis (and hence
the center of the ponderomotive potential well), we use pulses such
that $\epsilon_{\textrm{d}}\ll\xi_{0}^{-1}\ll1$. The electromagnetic
fields are obtained via the identities \cite{Jackson1975}
\begin{equation}
\begin{array}{c}
\vec{B}=\nabla\times\vec{A}\\
\vec{E}=\textrm{c}^{2}\nabla\int\nabla\cdot\vec{A}\textrm{d}t-\dfrac{\partial\vec{A}}{\partial t}
\end{array},\label{EM}
\end{equation}
in which we have applied the Lorenz gauge. 

Consider a non-zero $\theta$ ($\theta=\theta_{\textrm{t}}$ or $\theta_{\textrm{l}}$)
and a particle propagating in the +z direction with speed $\left|\vec{v}\right|\equiv\beta\textnormal{c}$.
We henceforth denote all variables in the native frame of the electromagnetic
pulse with prime superscripts, so the pulse propagates in the $+\textnormal{z}'$
direction, and all variables in the particle's rest frame with double-prime
superscripts. Non-primed variables $x,y,z,t$ are lab frame variables,
defined in accordance with Fig. \ref{Fig:Schematic}(a). Note that in the rest frame, $\omega$
in (\ref{Up}) should be replaced by the Doppler-shifted frequency $\omega''\equiv\omega\gamma\left(1-\beta\cos\theta\right)$.
Applying the appropriate rotation and Lorentz transformation operators
to (\ref{A}) and (\ref{EM}), we obtain the ponderomotive potential
in the rest frame as
\begin{eqnarray}
U_{\textrm{p}}'' & = & \frac{q^{2}}{4m_{0}}\left(\left|\tilde{A_{x}'}\right|^{2}+\left|\tilde{A_{y}'}\right|^{2}\right)g^{2}\left[1+\textnormal{O}\left(\epsilon_{\textnormal{d}}\right)+\textnormal{O}\left(\xi_{0}^{-1}\right)+\textnormal{O}\left(\beta\right)\right],\label{Up_prpr}
\end{eqnarray}
a result that applies for general $\vec{\tilde{A}}$ satisfying the
paraxial wave equation, assuming that $\tilde{A_{\textnormal{z}}'}$
is on the order of the transverse components or less.

For the linearly-polarized Hermite Gaussian $\textnormal{TEM}_{mn}$
mode,
\begin{equation}
\vec{\tilde{A}}'\equiv\hat{\textnormal{x}}'A_{0}f'\exp\left(-f'\rho'^{2}\right)\textrm{H}_{m}\left(\left|f'\right|\tilde{x}'\right)\textrm{H}_{n}\left(\left|f'\right|\tilde{y}'\right)\left(\frac{f'}{f'^{*}}\right){}^{\frac{m+n}{2}},\label{A_HG}
\end{equation}
where $A_{0}$ is a normalization constant, $f'\equiv\textrm{i}/\left(\textrm{i}+z'/z_{0}\right)$,
$\tilde{x}'\equiv\sqrt{2}x'/w_{0}$, $\tilde{y}'\equiv\sqrt{2}y'/w_{0}$,
$z_{0}\equiv\textrm{\ensuremath{\pi}}w_{0}^{2}/\lambda$ is the Rayleigh
range, $w_{0}$ is the beam waist radius, $\rho'\equiv\sqrt{x'^{2}+y'^{2}}/w_{0}$,
and $\textrm{H}_{m}\left(\cdot\right)$ is the Hermite polynomial
of order $m$ ($\textrm{H}_{0}\left(x\right)=1$, $\textrm{H}_{1}\left(x\right)=2x$
etc.), with $m,n\in\mathbb{N}_{0}$ (the set of natural numbers including
0). The beam divergence angle is $\epsilon_{\textnormal{d}}\equiv2/(kw_{0})$.
From (\ref{A}) and (\ref{EM}), the peak power $P$ transported in
the propagation direction is
\begin{equation}
P\equiv\iint S_{\textrm{z}0}'\textrm{d}x'\textrm{d}y'\thickapprox\omega^{2}A_{0}^{2}\textnormal{c}\epsilon_{0}\pi w_{0}^{2}2^{n+m-1}n!m!,\label{eq:P}
\end{equation}
where $S_{\textrm{z}0}'$ denotes the z-directed Poynting vector $S_{\textrm{z}}'\equiv\vec{E}'\times\vec{H}'\cdot\hat{z}'$
evaluated at the pulse peak, focal plane and carrier amplitude. $\epsilon_{0}$
is the permittivity of free space. The energy $U$ of a single pulse
is related to its peak power as
\begin{equation}
U\equiv\iiint S_{\textrm{z}}'|{}_{z'=0}\textrm{d}x'\textrm{d}y'\textrm{d}t'\thickapprox\frac{P}{2}\int g^{2}\left(\frac{\xi'}{\xi_{0}}\right)\textrm{d}\left(\frac{\xi'}{\omega}\right).\label{eq:U}
\end{equation}

Longitudinal compression is achieved with the $\textnormal{TEM}_{mn}$
mode when $m$ is odd and $n$ is even. In that case,
\begin{eqnarray}
U_{\textrm{pl}}'' & = & \frac{m_{0}K_{\textnormal{l}}}{2}\left[\int g^{2}\left(\frac{\xi'}{\xi_{0}}\right)\textrm{d}\left(\frac{\xi'}{\omega}\right)\right]^{-1}x'^{2}g^{2}\left(\frac{\xi'}{\xi_{0}}\right)\nonumber \\
 &  & \left[1+\textnormal{O}\left(\xi_{0}^{-1}\right)+\textnormal{O}\left(\left(n+m+1\right)\epsilon_{\textnormal{d}}^{2}\right)+\textnormal{O}\left(\beta\right)\right],\label{Upl}
\end{eqnarray}
where
\begin{equation}
K_{\textnormal{l}}\equiv\frac{q^{2}\lambda^{2}}{\pi^{3}m_{0}^{2}\epsilon_{0}\textnormal{c}^{3}}\frac{U}{w_{0}^{4}}\frac{m!n!}{2^{m+n-2}\left[\left(m-1\right)/2\right]!^{2}\left(n/2\right)!^{2}},\label{Cl}
\end{equation}
and we have applied Taylor expansions about the origin in (\ref{Up_prpr})
to obtain (\ref{Upl}). The net impulse in the rest frame is then
\begin{eqnarray}
\triangle\vec{p_{\textnormal{l}}}'' & = & -\int\nabla''U_{\textrm{p}}''\textrm{d}t''\nonumber \\
 & = & m_{0}K_{\textnormal{l}}\frac{\left[\gamma\left(\beta-\cos\theta\right)\Delta x''+\sin\theta\Delta z''\right]}{\gamma^{2}\left(1-\beta\cos\theta\right)^{3}}\left[\textnormal{\ensuremath{\hat{x}}}\left(\cos\theta-\beta\right)-\textnormal{\ensuremath{\hat{z}}}\frac{\sin\theta}{\gamma}\right]\nonumber \\
 &  & \left[1+\textnormal{O}\left(\xi_{0}^{-1}\right)+\textnormal{O}\left(\left(n+m+1\right)\epsilon_{\textnormal{d}}^{2}\right)+\textnormal{O}\left(\beta\right)\right],\label{dp_prpr_l}
\end{eqnarray}
where the particle's rest frame displacement from the bunch centroid
is $\left(\Delta x'',\Delta y'',\Delta z''\right)$, which we assume
does not change significantly during the interaction. To eliminate
the x-directed modulation and the $\Delta x''$-dependence of the
z-directed modulation in the lowest-order term, we must choose $\theta$
such that $\cos\theta=\beta$, a condition equivalent to (\ref{best theta}).
The lab-frame velocity change is then
\begin{eqnarray}
\triangle\vec{v_{\textnormal{l}}} & = & -\hat{z}K_{\textnormal{l}}\Delta z\left[1+\textnormal{O}\left(\xi_{0}^{-1}\right)+\textnormal{O}\left(\left(n+m+1\right)\epsilon_{\textnormal{d}}^{2}\right)+\textnormal{O}\left(\beta\right)\right],\label{dv_l}
\end{eqnarray}
where the particle's lab frame displacement from the bunch centroid
is $\left(\Delta x,\Delta y,\Delta z\right)$. The longitudinal impulse
in the lab frame follows from the relation $\triangle\vec{p_{\textnormal{l}}}=m_{0}\gamma^{3}\triangle\vec{v_{\textnormal{l}}}+\textnormal{O\ensuremath{\left(\triangle v_{\textnormal{l}}^{2}\right)}}$.
The linear dependence in the lowest-order term of (\ref{dv_l}) corresponds
to a parabolic potential profile. In the absence of space-charge and
momentum spread, a particle pulse would be compressed by a perfectly
parabolic potential to a zero extent.

Transverse compression is achieved with the $\textnormal{TEM}_{mn}$
mode when $m$ is even and $n$ is odd. In this case,
\begin{eqnarray}
U_{\textrm{pt}}'' & = & \frac{m_{0}K_{\textnormal{t}}}{2}\left[\int g^{2}\left(\frac{\xi'}{\xi_{0}}\right)\textrm{d}t'\right]^{-1}y'^{2}g^{2}\left(\frac{\xi'}{\xi_{0}}\right)\nonumber \\
 &  & \left[1+\textnormal{O}\left(\xi_{0}^{-1}\right)+\textnormal{O}\left(\left(n+m+1\right)\epsilon_{\textnormal{d}}^{2}\right)+\textnormal{O}\left(\beta\right)\right],\label{Upt}
\end{eqnarray}
where
\begin{equation}
K_{\textnormal{t}}\equiv\frac{q^{2}\lambda^{2}}{\pi^{3}m_{0}^{2}\epsilon_{0}\textnormal{c}^{3}}\frac{U}{w_{0}^{4}}\frac{m!n!}{2^{m+n-2}\left[\left(n-1\right)/2\right]!^{2}\left(m/2\right)!^{2}}.\label{Ct}
\end{equation}

The net transverse impulse imparted by the passage of a single pulse
in the rest frame is
\begin{eqnarray}
\triangle\vec{p_{\textnormal{t}}}'' & = & -m_{0}K_{\textnormal{t}}\frac{1}{\gamma\left(1-\beta\cos\theta\right)}\Delta y''\textnormal{\ensuremath{\hat{y}}}\nonumber \\
 &  & \left[1+\textnormal{O}\left(\xi_{0}^{-1}\right)+\textnormal{O}\left(\left(n+m+1\right)\epsilon_{\textnormal{d}}^{2}\right)+\textnormal{O}\left(\beta\right)\right],\label{dp_prpr_t}
\end{eqnarray}
corresponding to a net lab-frame velocity change of 
\begin{eqnarray}
\triangle\vec{v_{\textnormal{t}}} & = & -\hat{y}K_{\textnormal{t}}\frac{1}{\gamma^{2}\left(1-\beta\cos\theta\right)}\Delta y\nonumber \\
 &  & \left[1+\textnormal{O}\left(\xi_{0}^{-1}\right)+\textnormal{O}\left(\left(n+m+1\right)\epsilon_{\textnormal{d}}^{2}\right)+\textnormal{O}\left(\beta\right)\right].\label{dv_t}
\end{eqnarray}
As $\theta$ approaches $0^{\textnormal{o}}$, the velocity change
becomes larger a result of improved group velocity matching along
the optical beam axis. The transverse impulse in the lab frame follows
from the relation $\triangle\vec{p_{\textnormal{t}}}=m_{0}\gamma\triangle\vec{v_{\textnormal{t}}}+\textnormal{O\ensuremath{\left(\triangle v_{\textnormal{t}}^{2}\right)}}$.
Several noteworthy features of the pulse compression scheme are evident
from (\ref{Cl}), (\ref{dv_l}), (\ref{Ct}) and (\ref{dv_t}):

1. At the lowest order, net velocity change is independent of pulse
duration parameter $\xi_{0}$ and pulse shape $g$.

2. A trade-off between the size of the parabolic potential region and
the strength of the compression exists in two ways: through the laser
waist radius $w_{0}$, and through the choice of $m$ and $n$. One
solution to achieving a large parabolic potential region and a large
$\triangle v$ for a given total optical energy may lie in the superposition
of higher-order Hermite-Gaussian modes, as proposed in \cite{Steuernagel2009}
in the context of atomic beam imaging.

3. That $\triangle v\propto\lambda^{2}$ (as expected of a ponderomotive force scheme \cite{Boot1957}) suggests that greater net impulse may be achieved via longer-wavelength sources. Note, however, that increasing the wavelength increases the pulse duration for the same
number of temporal cycles, which may weaken the assumption that the
particle's position relative to the intensity well does not change
significantly during the interaction.

The focal time (the time of maximal compression) of a particle pulse
with a velocity chirp can be estimated with the formula
\begin{equation}
t_{\textnormal{f}}=\frac{\triangle r_{0}}{\left|v_{\textnormal{T}}\right|},\label{t_f_analy}
\end{equation}
where $v_{\textnormal{T}}\equiv\triangle v+v_{0}$. $\triangle r_{0}$ 
and $v_{0}$  refer respectively to the half-width of the particle pulse and the velocity of a
particle at the pulse's edge, along the dimension of compression and immediately before the interaction. 
$\triangle v$ is the velocity change induced in the particle at the pulse's edge as a result of the 
interaction.

\section{ Numerical simulations}

To numerically model the laser-electron interaction, we solve the
exact Newton-Lorentz equation using an adaptive-step fifth-order Runge-Kutta
algorithm \cite{Press1992}. The coordinates of each particle are
assigned in a quasi-random fashion using Halton sequences \cite{Press1992}.
For the laser pulses, we employ first-order Hermite-Gaussian modes
that are exact (i.e. non-paraxial) solutions of Maxwell's equations
in free space. We readily obtain the fields of a TEM\textsubscript{10}
mode with a Poisson spectrum by using the Hertz vector potential
\begin{equation}
\vec{\Pi}_{10}=\frac{\textnormal{\ensuremath{\partial}}}{\textnormal{\ensuremath{\partial}}x}\vec{\Pi}_{\textnormal{00}}\label{Hertz pot}
\end{equation}
in the relations \cite{Sheppard2000}
\begin{equation}
\vec{B} =  \textrm{Re}\left\{ \frac{1}{\textnormal{c}^{2}}\frac{\partial}{\partial t}\nabla\times\vec{\Pi}_{\textnormal{10}}\right\} \nonumber 
\end{equation}
\begin{equation}
\vec{E} =  \textrm{Re}\left\{ \nabla\times\nabla\times\vec{\Pi}_{\textnormal{10}}\right\}.\label{Hertz to EM}
\end{equation}
The vector potential corresponding to a fundamental Gaussian
mode is \cite{April2010,Wong2014}

\begin{equation}
\vec{\Pi}_{\textnormal{00}}=\hat{\textnormal{x}}\Pi_{0}\frac{1}{R'}\left(f_{+}^{-s-1}-f_{-}^{-s-1}\right),\label{00 Hertz pot}
\end{equation}
where $f_{\pm}=1-\left(\textnormal{i}/s\right)\left(\omega t\pm kR'+\textnormal{i}ka\right)$,
$R'=\left[x^{2}+y^{2}+\left(z+\textnormal{i}a\right)^{2}\right]^{1/2}$,
and $\Pi_{0}$ is a complex constant. The degree of focusing and the
pulse duration are controlled through parameters $a$ and $s$ via
relations for which good analytical approximations have been derived
\cite{Wong2014,April2010}. The non-paraxial Gaussian beam reduces
to the phasor of the paraxial Gaussian beam in the paraxial limit
\cite{April2008}, so the description (\ref{Hertz pot})-(\ref{00 Hertz pot})
is consistent with (\ref{A})-(\ref{A_HG}).

Unless otherwise specified, all numerical simulations use optical
pulses of wavelength $\lambda=0.8\textnormal{$\mu$m}$, waist
radius $w_{0}=180\textnormal{$\mu$m}$, and (intensity) full-width-half-maximum
(FWHM) pulse duration $\tau=50\,\textnormal{fs}$. Each optical pulse
in the longitudinal compression stage has an energy of 17.5 mJ, whereas
each pulse in the transverse compression stage has an energy of about
26 mJ. Such specifications fall well within the realm of what is experimentally
achievable today. The initial 30 keV electron pulse is a zero-emittance,
uniformly-filled ellipsoid of diameter 28 $\mu$m and length 14 $\mu$m, corresponding
to a FWHM electron pulse duration of 100 fs. The particles are non-interacting
and our simulation results are thus applicable to single-electron
pulses. Although actual electron pulses have non-zero emittances that
vary depending on factors like the the type of emission mechanism
used \cite{Carbone2012,Baum2013}, we use electron pulses with zero
initial emittance to perform numerical evaluations of our scheme that
are independent of non-idealities in the initial electron pulse.

\begin{figure}
\includegraphics[width=16cm]{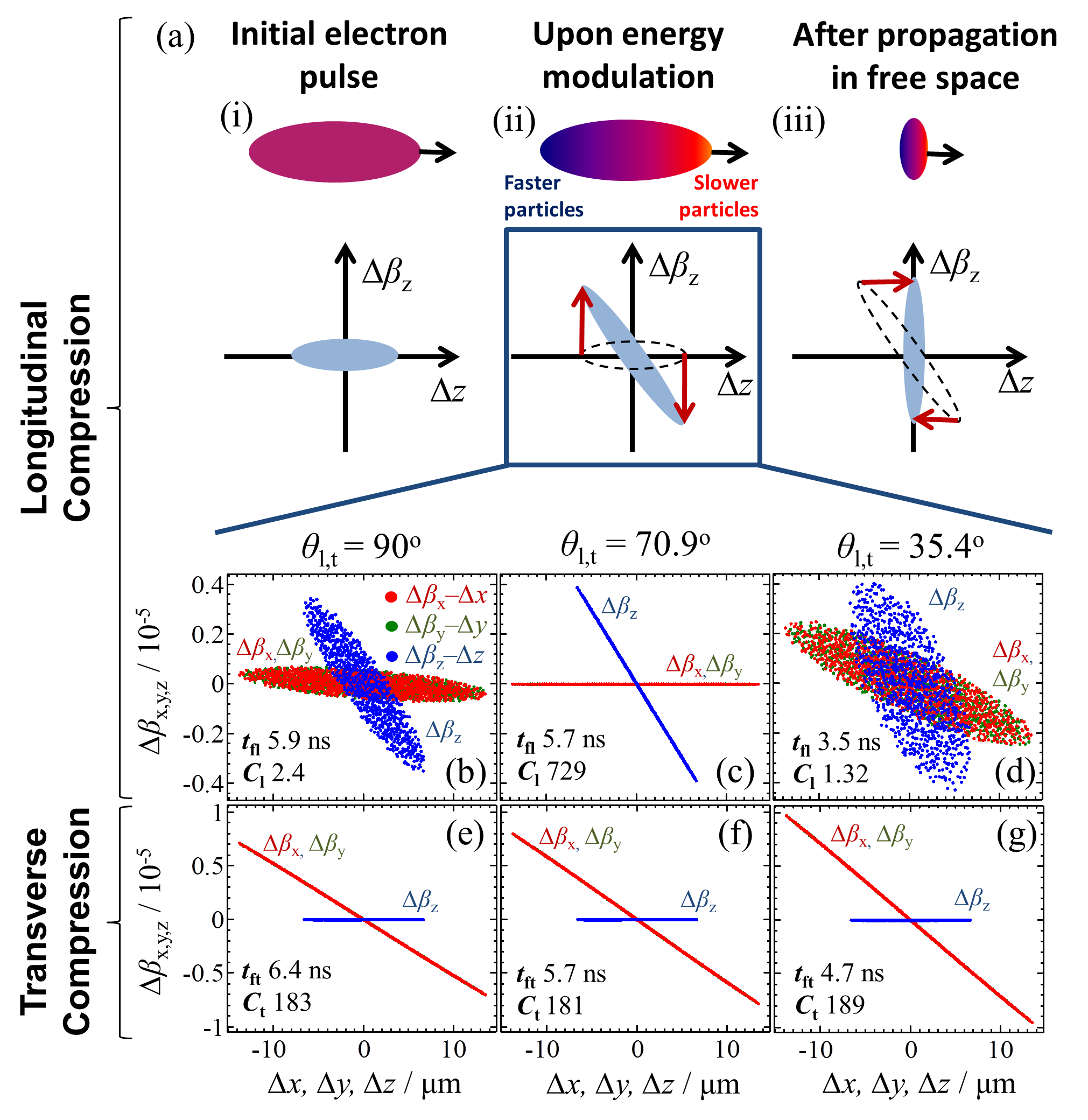}\caption{(a) Physical mechanism of the longitudinal compression scheme: (i) the initial electron pulse has a finite spread in momentum and position; (ii) the laser-electron interaction accelerates the back electrons and decelerates the front electrons; (iii) As the pulse propagates, the back electrons catch up with the front electrons, leading to electron pulse compression.  $\Delta z \equiv z-\left\langle z\right\rangle $ denotes the particle’s displacement from the bunch centroid along the z-dimension (and so on for the other variables). Phase plane distributions of the 30 keV electron pulse immediately after
the longitudinal compression stage are shown for optical incidence angles (b) $\theta_{\textnormal{l}}=90^{\textnormal{o}}$,
(c) $\theta_{\textnormal{l}}=\arctan\left(1/\gamma\beta\right)=70.9^{\textnormal{o}}$, 
 and (d) $\theta_{\textnormal{l}}=35.4^{\textnormal{o}}$. Phase plane distributions of the electron pulse immediately after the transverse compression stage are shown for optical incidence angles (e) $\theta_{\textnormal{t}}=90^{\textnormal{o}}$,
(f) $\theta_{\textnormal{t}}=70.9^{\textnormal{o}}$and
(g) $\theta_{\textnormal{t}}=35.4^{\textnormal{o}}$. 
Focal times $t_\textnormal{fl,ft}$ and compression ratios $C_\textnormal{l,t}$ are indicated for each case.
1000 particles were used in each simulation. \label{Fig:DifferentExamples}}
\end{figure}

Figs. \ref{Fig:DifferentExamples}(b)-(d) depict the numerically-computed
phase space distributions of electron pulses immediately after the
longitudinal compression stage, for various optical incidence angles
$\theta_{\textnormal{l}}$. The longitudinal magnification $M_{\textnormal{l}}$
is defined as $M_{\textnormal{l}}\equiv\sigma_{\textnormal{z}}\left(t_{\textnormal{fl}}\right)/\sigma_{\textnormal{z}}\left(0\right)$,
where $\sigma_{\textnormal{z}}=\sigma_{\textnormal{z}}\left(t\right)$
is the standard deviation in z at time $t$. Here, $t=0$ is defined
as the instant captured in Fig. \ref{Fig:DifferentExamples} (a)(ii) and $t=t_{\textnormal{fl}}$
the instant when the longitudinal focus is achieved (i.e. when
$M_{\textnormal{l}}$ is minimized, captured in Fig. \ref{Fig:DifferentExamples} (a)(iii)). The transverse magnification at
the longitudinal focus is $M_{\textnormal{tl}}\equiv\sigma_{\textnormal{x}}\left(t_{\textnormal{fl}}\right)/\sigma_{\textnormal{x}}\left(0\right),$
where $\sigma_{\textnormal{x}}=\sigma_{\textnormal{x}}\left(t\right)$
is the standard deviation in x at time $t$. In Fig. \ref{Fig:DifferentExamples}(b),
we see two undesirable effects of normal optical incidence in the
lab frame, both as analytically predicted in (\ref{dp_prpr_l}): the
significant modulation in the transverse phase planes, and the substantial
smear in the $\triangle\beta_{\textnormal{z}}$-$\triangle z$ phase
plane, resulting in a large longitudinal emittance and consequently
a weak longitudinal compression factor $C_{\textnormal{l}}\equiv M_{\textnormal{l}}^{-1}$.
The smeared particle distributions are largely due to walk-off between
the center of the ponderomotive potential well and the center of the
electron pulse, whereas the presence of transverse modulation is largely
due to the oblique optical incidence angle in the rest frame of the
electron pulse.

Note that the smearing and transverse modulation exist in spite of
the fact that the optical pulse duration $\tau=50$ fs is several tens of times
smaller than $w_{0}/v$ ($w_{0}\left(v\tau\right)^{-1}\approx36\gg1$),
and so nominally satisfies the thin lens approximation condition prescribed
in \cite{Hilbert2009} for normal incidence. This suggests that the
thin lens approximation condition alone is not sufficient for effective
longitudinal compression when the kinetic energy is on the order of
30 keV or greater. 

\begin{figure}
\includegraphics[width=16cm]{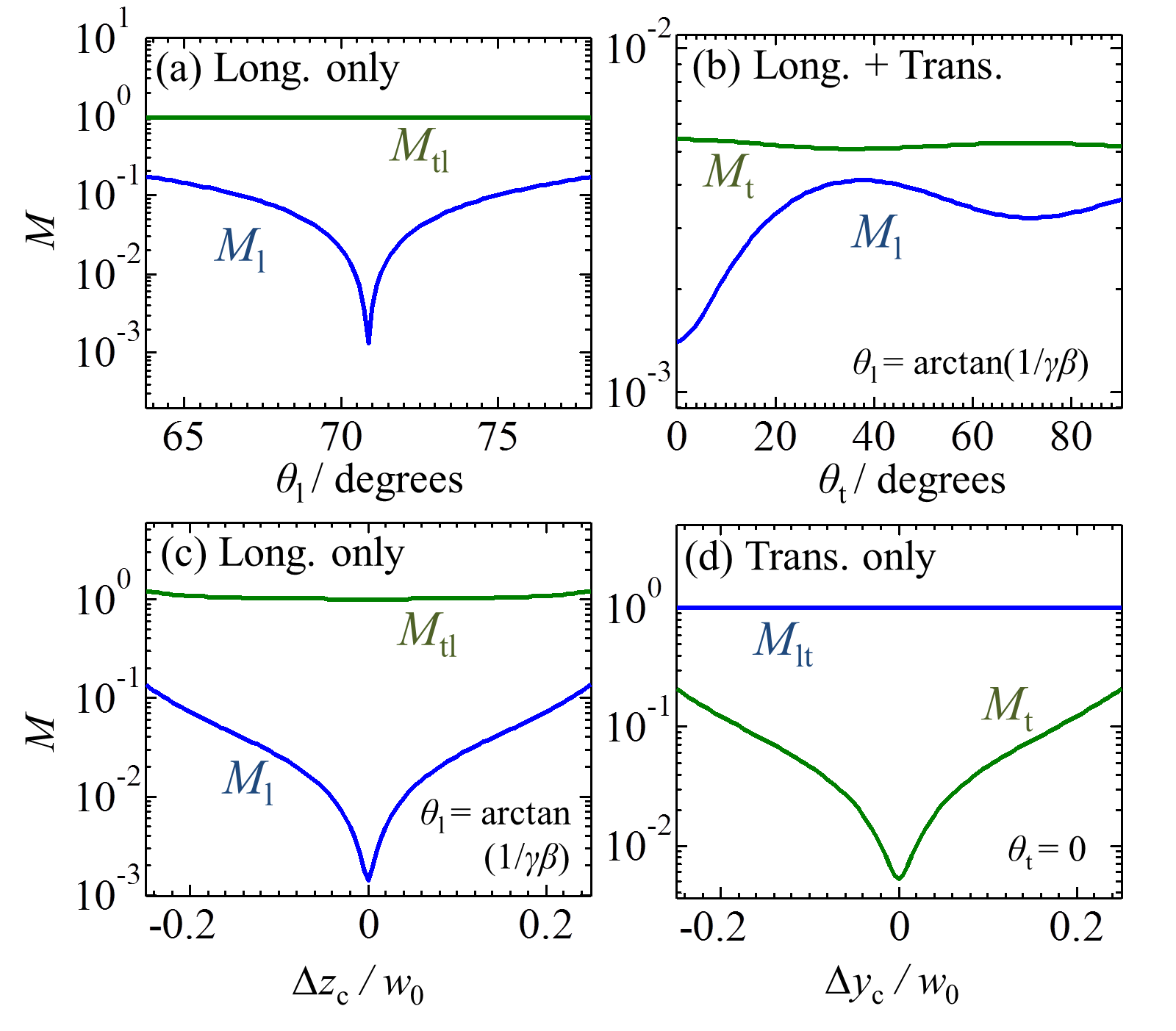}\caption{Sensitivity of compression ratios to (a) $\theta_{\textnormal{l}}$
in a longitudinal compression scheme, (b) $\theta_{\textnormal{t}}$ in a three-dimensional compression scheme, 
(c) longitudinal displacement $\Delta$$z_{\textnormal{c}}$ of the bunch centroid from the intensity minimum in a longitudinal compression
scheme, and (d)  transverse displacement  $\Delta$$y_{\textnormal{c}}$ of the bunch centroid from the intensity minimum in a transverse compression
scheme (along y). \label{Fig:AngleSensitiv}}
\end{figure}

As (\ref{dp_prpr_l}) predicts, injecting the optical pulse at an
oblique angle according to (\ref{best theta}) decouples the longitudinal
modulation from the transverse modulation at the lowest order and
significantly improves the compression factor from the normal incidence
case in Fig. \ref{Fig:DifferentExamples}(b). This is shown in Fig.
\ref{Fig:DifferentExamples}(c), where we achieve a compression factor
of $C_{\textnormal{l}}=729$, taking the 100 fs electron pulse well
into the attosecond regime. Further decreasing the incidence angle,
as we do in Fig. \ref{Fig:DifferentExamples}(d), gives rise again
to the substantial smearing of particle distributions in the $\triangle\beta_{\textnormal{z}}$-$\triangle z$
phase plane, as well as modulations in the transverse phase planes. The sensitivity of the longitudinal compression to
the optical incidence angle in the lab frame is further illustrated in Fig. \ref{Fig:AngleSensitiv}(a).

Note that the area occupied in a 2-dimensional phase plane is not
conserved in the interaction due to inter-dimensional coupling caused
by a non-zero magnetic field. This does not violate Liouville's theorem,
which states that the 6-dimensional phase space volume is conserved
in a Hamiltonian system. Note also that the electron pulse is affected
equally in the $\triangle\beta_{\textnormal{x}}$-$\triangle x$ and
$\triangle\beta_{\textnormal{y}}$-$\triangle y$ phase planes due
to our use of both Pulses I and II in Fig. \ref{Fig:Schematic}(a),
instead of attempting the longitudinal compression with only one of
them.

Figs. \ref{Fig:DifferentExamples}(e)-(g) depict the numerically-computed
phase space distributions of electron pulses immediately after the
transverse compression stage, for various optical incidence angles
$\theta_{\textnormal{t}}$. The transverse magnification is defined
as $M_{\textnormal{t}}\equiv\sigma_{\textnormal{x}}\left(t_{\textnormal{ft}}\right)/\sigma_{\textnormal{x}}\left(0\right),$
where $t_{\textnormal{ft}}$ is the time at which $M_{\textnormal{t}}$
is minimal. The longitudinal magnification at the transverse focus
is $M_{\textnormal{lt}}\equiv\sigma_{\textnormal{z}}\left(t_{\textnormal{ft}}\right)/\sigma_{\textnormal{z}}\left(0\right)$.
Note that because the configuration in Fig. \ref{Fig:Schematic}(a)
subjects the electron pulse to similar treatments in x and y at the
lowest order, $\sigma_{\textnormal{y}}$ behaves essentially in the
same way as $\sigma_{\textnormal{x}}$. The increase in $\triangle\beta_{\textnormal{x,y}}$
(and subsequent decrease in $t_{\textnormal{ft}}$) as $\theta_{\textnormal{t}}$
decreases is as analytically predicted in (\ref{dv_t}). Although
the transverse compression ratio is a relatively weak function of
$\theta_{\textnormal{t}}$, we see in Fig. \ref{Fig:AngleSensitiv}(b)
that the choice of $\theta_{\textnormal{t}}$ can significantly affect
the longitudinal compression ratio in a three-dimensional compression
scheme via higher-order terms of the transverse compression stage,
with maximum longitudinal compression achieved when $\theta_{\textnormal{t}}=0^{\textnormal{o}}$. 

Figs. \ref{Fig:AngleSensitiv} (c) 
and (d) show the sensitivity of the compression to the displacement of the bunch centroid from the intensity minimum during the interaction. 
For the cases considered, one should aim to keep $\theta_{\textnormal{l}}$ within $0.1^{\textnormal{o}}$ of 
its optimal value, and the electron bunch centroid within $0.01w_{\textnormal{0}}$ of the intensity minimum (values approximated by considering the half-width-half-maximum of the compression).

Using the simulation parameters of Fig. \ref{Fig:DifferentExamples}(c) in (\ref{dv_l}),
we obtain $\triangle v_{\textnormal{l}}\approx3.925\times10^{-6}\textnormal{c}$
at $\triangle z=-6.627$ \textmu{}m (actual value $\triangle\beta_{\textnormal{z}}=3.894\times10^{-6}$,
relative error 0.80\%). For the transverse compression cases, (\ref{dv_t}) yields $\triangle v_{\textnormal{t}}\approx3.671\times10^{-6}\textnormal{c}$
at $\triangle x=-6.945$ \textmu{}m in Fig. \ref{Fig:DifferentExamples}(e)
(actual value $\triangle\beta_{\textnormal{x}}=3.644\times10^{-6}$,
relative error 0.74\%) ; $\triangle v_{\textnormal{t}}\approx4.113\times10^{-6}\textnormal{c}$
at $\triangle x=-6.944$ \textmu{}m in Fig. \ref{Fig:DifferentExamples}(f)
(actual value $\triangle\beta_{\textnormal{x}}=4.081\times10^{-6}$,
relative error 0.78 \%); and $\triangle v_{\textnormal{t}}\approx5.010\times10^{-6}\textnormal{c}$
at $\triangle x=-6.943$ \textmu{}m in Fig. \ref{Fig:DifferentExamples}(g)
(actual value $\triangle\beta_{\textnormal{x}}=4.974\times10^{-6}$,
relative error 0.72 \%). These examples demonstrate the accuracy of
(\ref{dv_l}) and (\ref{dv_t}) in estimating the velocity chirp induced
by the interaction.

While the momentum modulations in these examples are small, they are still more than two orders of magnitude greater than the minimum momentum spread required by the Heisenberg uncertainty principle for the electron pulse dimensions considered ($\Delta x\Delta p_{\textnormal{x}}\geq\hbar/2$ gives $\Delta\gamma\beta_x\geq1.4\times10^{-8}$ for $\Delta x = 14\ \mu \textnormal{m}$). Nevertheless, actually producing an electron bunch with an emittance small enough for the bunch to be affected by such small modulations is currently very challenging (required emittance $\epsilon_x \sim \Delta x\Delta\gamma\beta_{\textnormal{x}} <  0.1\  \textnormal{nm}$). In actual implementations, larger momentum modulations -- and hence more realistic emittance requirements -- are readily achievable by, for instance, increasing the laser intensity (e.g., by tighter focusing) or introducing more stages in the cascade of Fig. \ref{Fig:Schematic}(b). For single-electron sources in the tens-of-keV range, emittances as low as 3 nm have been demonstrated \cite{Baum2013}.

\begin{figure}
\includegraphics[width=16cm]{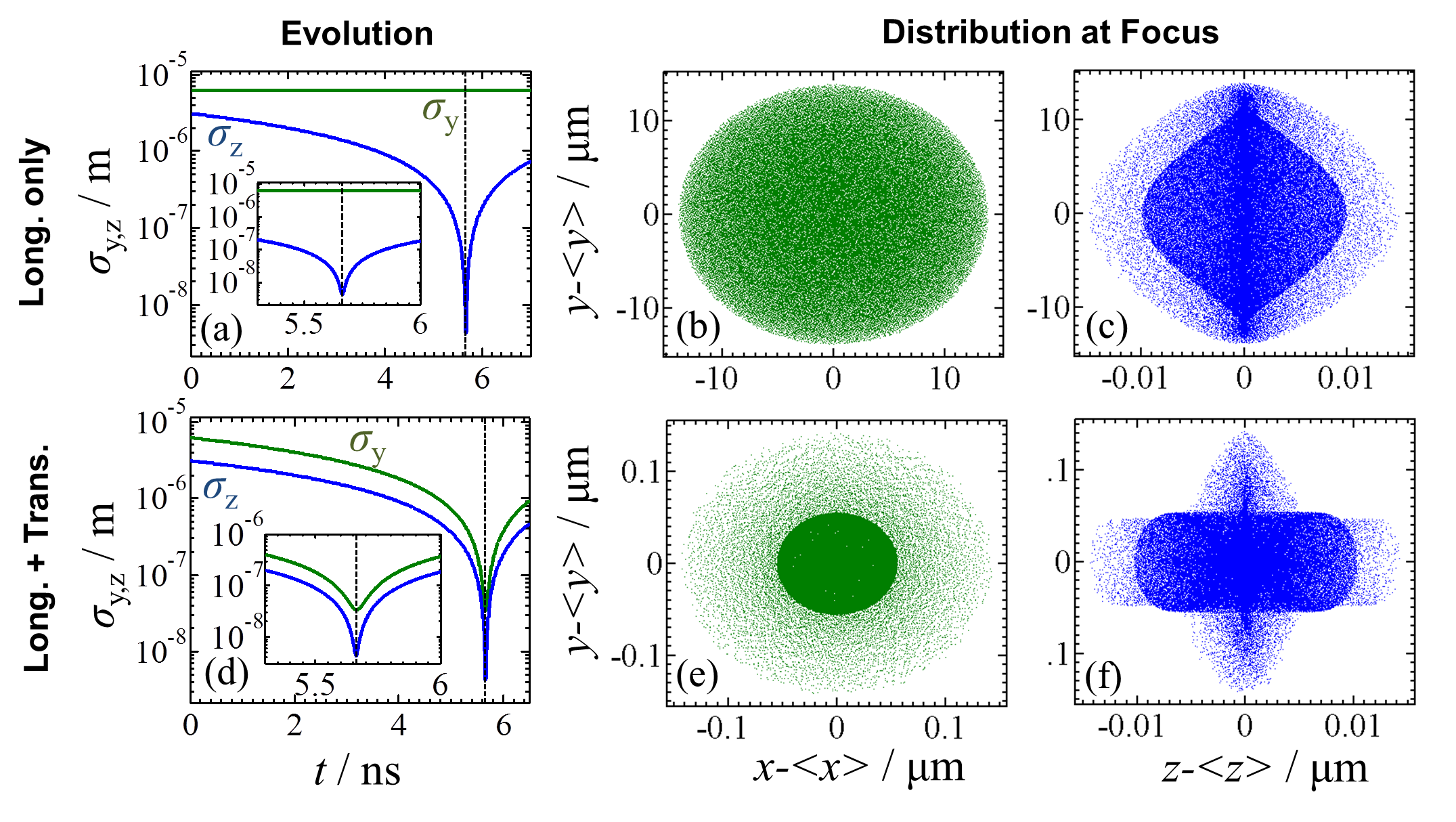}\caption{Longitudinal compression of a 30 keV electron pulse from 100 fs to
137 as (longitudinal compression factor $C_{\textnormal{l}}=729$):
(a) Evolution of standard deviations in y and z, and corresponding spatial distributions
at the longitudinal focus (vertical dotted line in (a))
in the (b) y-x and (c) y-z planes. Three-dimensional
compression of a 30 keV electron pulse from a duration of 100 fs and
a diameter of 28 $\mu$m to a duration of 137 as and an effective diameter
of 0.153 $\mu$m ($C_{\textnormal{l}}=729$, $C_{\textnormal{t}}=183$):
(d) Evolution of standard deviations in y and z, and corresponding spatial distributions
at the focus (vertical dotted line in (d)) in the (e)
y-x and (f) y-z planes. $10^{5}$ particles
were used in each simulation. In (a) and (d), the standard deviation
in x was omitted as it practically lies over that in y.\label{Fig:ExampleLongCompression}}
\end{figure}

The negative velocity chirp in Fig. \ref{Fig:DifferentExamples}(c)
causes the 30 keV electron pulse to compress longitudinally
as it continues propagating after the interaction. Fig. \ref{Fig:ExampleLongCompression}(a)
shows the evolution of the electron pulse's transverse and longitudinal
standard deviations with time. Note that the transverse spread remains
practically unchanged from its initial value, even as the electron
pulse is compressed from a pulse duration of 100 fs to one of 137 as ($C_{\textnormal{l}}=729$).
The electron pulse distribution at the longitudinal focus, marked
by a vertical dotted line in Fig. \ref{Fig:ExampleLongCompression}(a),
is shown in Figs. \ref{Fig:ExampleLongCompression}(b) and (c). The
higher-order non-linear components of the induced velocity chirp prevents
the ellipsoid from collapsing into a perfectly flat pancake. 

Figs. \ref{Fig:ExampleLongCompression}(d), (e) and (f) depict the
three-dimensional compression of a 30 keV electron pulse from a duration
of 100 fs and a diameter of 28 $\mu$m to a duration of 137 as and a diameter
of 0.153 $\mu$m ($C_{\textnormal{l}}=729$, $C_{\textnormal{t}}=183$).
$\theta_{\textnormal{l}}$ satisfies (\ref{best theta}) and $\theta_{\textnormal{t}}=0^{\textnormal{o}}$.
Note that simultaneous transverse and longitudinal compression is
achieved without affecting the longitudinal compression ratio of the
purely-longitudinal-compression scheme in Fig. \ref{Fig:ExampleLongCompression}(a).

\begin{figure}
\includegraphics[width=16cm]{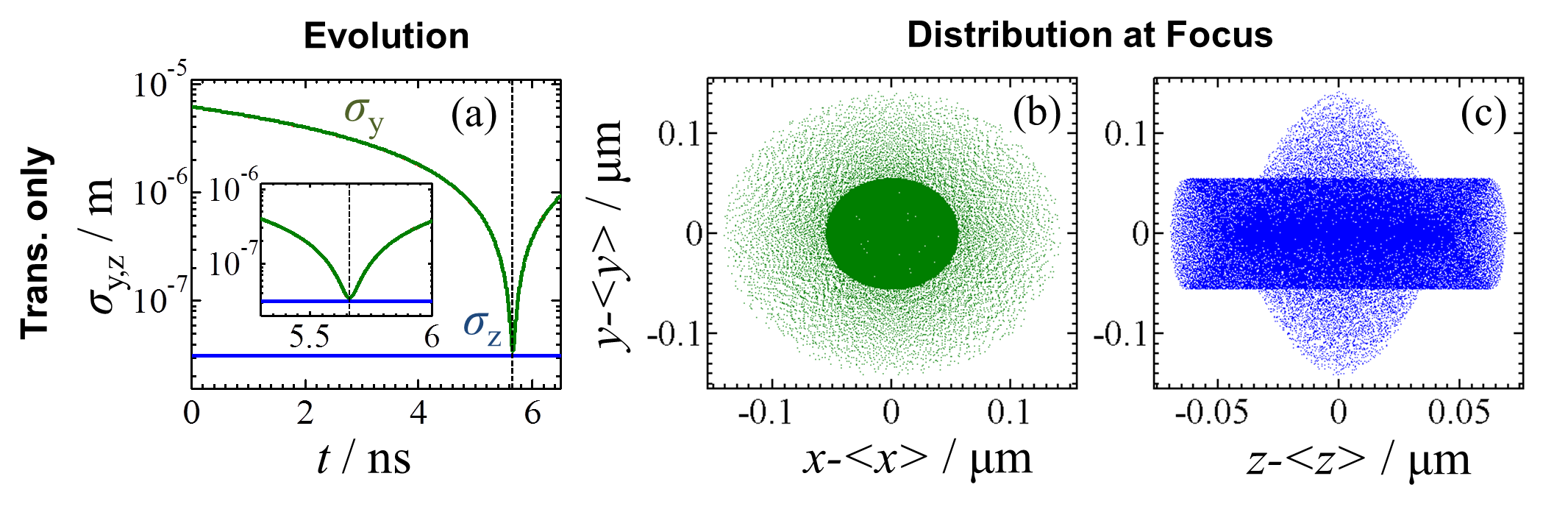}\caption{Transverse compression of a 30 keV, 1 fs-long electron pulse from a
diameter of 28 $\mu$m to an effective diameter of 0.156 $\mu$m (transverse
compression factor $C_{\textnormal{t}}=179$): (a) Evolution of standard
deviations in y and z, and corresponding spatial distributions at the transverse
focus (vertical dotted line in (a)) in the (b) y-x and
(c) y-z planes.\label{Fig:ExampleTxCompression}}
\end{figure}

Fig. \ref{Fig:ExampleTxCompression} depicts the transverse compression
of a 30 keV, 1 fs electron pulse from a diameter of 28 $\mu$m to one of 0.156 $\mu$m ($C_{\textnormal{t}}=179$). $\theta_{\textnormal{t}}=0^{\textnormal{o}}$
here. In Fig. \ref{Fig:ExampleTxCompression}(a), we see that the
longitudinal spread remains practically unchanged from its initial
value even as the electron pulse is focused transversely to a very small spot. This demonstrates
the ability of the proposed scheme to focus ultrashort electron pulses
without inducing temporal resolution-limiting distortions in them. 

\begin{figure}
\includegraphics[width=16cm]{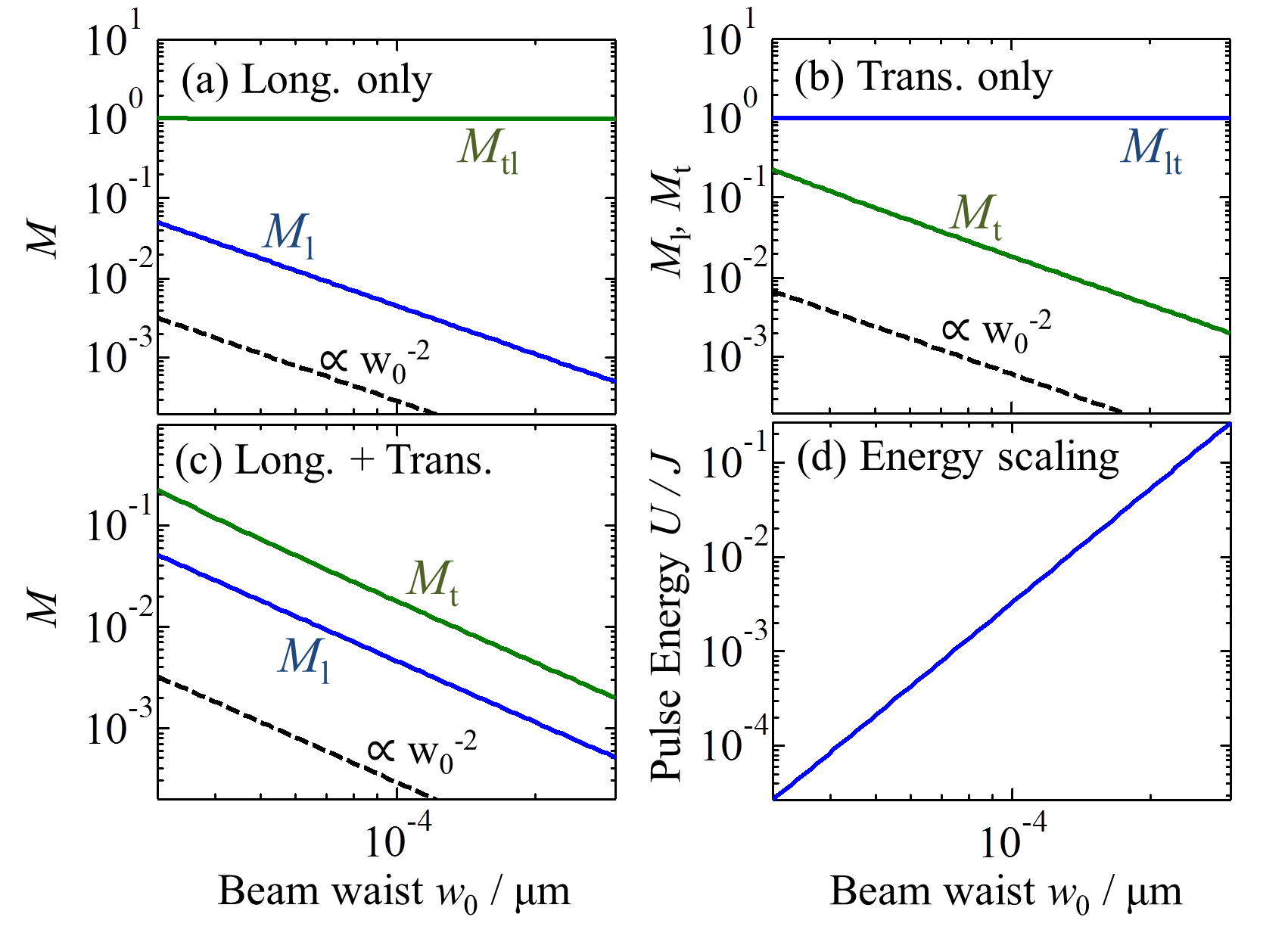}
\caption{Scaling of magnification with optical beam waist for the optical and electron pulses of Figs. \ref{Fig:DifferentExamples} and \ref{Fig:ExampleLongCompression}: (a) Longitudinal compression
only (b) Transverse compression only and (c) Three-dimensional compression.
(d) plots the scaling of energy $U$ with beam waist $w_{0}$ when
$Uw_{0}^{-4}$ is kept constant, from $U=$27 $\mu$J at $w_{0}=$30 $\mu$m
to $U=$270 mJ at $w_{0}=$300 $\mu$m. $\theta_{\textnormal{l}}$ satisfies
(\ref{best theta}) and $\theta_{\textnormal{t}}=0^{\textnormal{o}}$.\label{Fig:EnergyScaling}}
\end{figure}

The $Uw_{0}^{-4}$ dependence in (\ref{Cl}) and (\ref{Ct}) implies
that significant energy savings are possible with tighter focusing. Decreasing
the beam waist radius, however, enhances higher-order distortions that limit
the maximum achievable compression. Fig. \ref{Fig:EnergyScaling}
illustrates the tradeoff between compression factor and pulse energy
for the value of $Uw_{0}^{-4}$ and the electron pulse used in Figs. \ref{Fig:DifferentExamples}
and \ref{Fig:ExampleLongCompression}. That the magnification scales as $w_{0}^{-2}$
is consistent with the fact that the dominant higher-order distortions
scale as $\textnormal{O}\left(\epsilon_{\textnormal{d}}^{2}\right)$
in (\ref{dv_l}) and (\ref{dv_t}). In Fig. \ref{Fig:EnergyScaling},
$U$ refers to the total energy used in the longitudinal compression
stage. Since $\theta_{\textnormal{t}}=0^{\textnormal{o}}$, one may infer from (\ref{dv_t}) that 
the energy used for compression in each transverse dimension is typically smaller
by a factor of about $\left(1+\beta\right)$ so that the longitudinal and transverse foci coincide in the three-dimensional
compression scheme. Fig. \ref{Fig:EnergyScaling} shows
that decent compression factors are already attainable with relatively
low-energy optical pulses. In Fig. \ref{Fig:EnergyScaling}(a), for
instance, a longitudinal compression factor of 20 is already achievable
with optical pulses of waist radius 30 $\mu$m and total energy 27 $\mu$J.

Although we have focused on single-electron pulses in this work, the
proposed scheme can also be used for multi-electron pulse compression.
This is especially (but not only) true when the electron pulse approximates
a uniformly-filled ellipsoid or contains a linear velocity chirp.
It should be noted, however, that typical multi-electron pulses have
much larger diameters -- which are on the order of a few hundreds
of $\mu$m -- than the electron pulses considered here, necessitating more
energetic optical pulses to achieve the same compression qualities and
focal times.

\section{Conclusion}

We have proposed an all-optical three-dimensional electron pulse compression
scheme. The scheme comprises a succession of Hermite-Gaussian optical
modes that effectively fashions a three-dimensional optical trap in
the electron pulse's rest frame. Compression in each Cartesian dimension
can be controlled without affecting electron pulse properties in orthogonal
dimensions, at the lowest order. We showed mathematically
that the right choice of optical incidence angle is necessary in longitudinal
compression so that the induced velocity change is not a function
of transverse coordinates and not accompanied by transverse phase
plane modulations, at the lowest order. Although the transverse compression ratio is a relatively
weak function of $\theta_{\textnormal{t}}$ for 30 keV electrons,
the choice of $\theta_{\textnormal{t}}$ can significantly affect
the longitudinal compression ratio in a three-dimensional compression
scheme, with maximum longitudinal compression achieved when $\theta_{\textnormal{t}}=0^{\textnormal{o}}$.
We also derived analytical
expressions approximating the net velocity change induced in a charged
particle by a Hermite-Gaussian optical pulse of arbitrary order and
incidence angle. These analytical expressions can be used to estimate the velocity
chirp aquired by an electron pulse as a result of the laser-electron interaction.

Finally, using optical pulses that are realizable experimentally,
we numerically demonstrated the longitudinal compression of a 30 keV
electron pulse from 100 fs to 137 as (729 times compression), the
three-dimensional compression of a 30 keV electron pulse from a duration
of 100 fs and a diameter of 28 $\mu$m to a duration of 137 as (729 times
compression) and a diameter of 0.153 $\mu$m (183 times compression) ,
and the transverse compression of a 1 fs long, 30 keV electron pulse
from a diameter of 28 $\mu$m to one of 0.156 $\mu$m (179 times compression).
Even larger compression factors are potentially possible with larger
beam waists, at the cost of focal time for a given optical pulse energy.
Our energy scaling studies show that a compression factor of 20 is
already achievable with a 27 $\mu$J optical pulse of waist radius 30 $\mu$m.
The required pulse energies may be lowered further still with the
cascade scheme of Fig. \ref{Fig:Schematic}(b).

The proposed scheme is useful in
ultrafast electron imaging for both single- and multi-electron pulse
compression, and as a means of focusing ultrashort
electron pulses without inducing temporal resolution-limiting
distortions in them. Broader applications of the mechanism studied here potentially include
compressing or focusing accelerated protons \cite{Robson2007} and
neutral atoms \cite{Eichmann2009}, enhancing the quantum degeneracy of electron packets \cite{Spence1994}, creating flat electron beams \cite{Zhu2014}, and creating ultrashort electron bunches for coherent terahertz emission \cite{Li2008}.

\ack{This work was partially supported by the U.S. Army Research Office Contract No. W911NF-13-D-0001, and the DARPA Young Faculty Award Grant No. D13AP00050. L.J.W. acknowledges support from the Agency for Science, Technology and Research, Singapore. T. R. acknowledges support from the Alexander von Humboldt Foundation, Germany.}
\newline

\section*{References}
\bibliographystyle{unsrt}
\bibliography{all-optical}
\clearpage

\end{document}